\pdfoutput=1
\documentclass[11pt]{article}
\usepackage[margin=1in]{geometry}
\usepackage{amsmath,amssymb,amsthm,booktabs,graphicx,natbib,hyperref}
\hypersetup{colorlinks=true,linkcolor=black,citecolor=black,urlcolor=blue}
\newtheorem{proposition}{Proposition}
\newtheorem{corollary}{Corollary}
\theoremstyle{remark}\newtheorem{remark}{Remark}
\newcommand{\D}{\mathcal{D}}\newcommand{\N}{\mathcal{N}}
\title{Maximum Scores and Treatment Feedback in Regression Discontinuity Designs}
\author{Subir Hait\\[2pt]
\small Department of Counseling, Educational Psychology, and Special Education\\
\small Michigan State University, East Lansing, MI 48824, USA\\
\small \texttt{haitsubi@msu.edu} \quad ORCID: 0009-0004-9871-9677}
\date{November 2026}
\begin{document}
\maketitle
\begin{abstract}
Hospitals use repeated risk scores to decide when to send an alert. Some evaluations summarize these scores by their maximum and use the maximum as the running variable in a regression discontinuity (RD) design. The maximum shows correctly whether a patient was ever alerted. However, it can include scores recorded after an earlier alert has already changed care. In this paper, we study what the maximum-score RD estimates when an earlier intervention changes a later score. We use a two-assessment model and derive a formula that splits the RD contrast into the causal effect and a composition term. When feedback moves scores in one direction, the composition term equals the relative jump in the density of the maximum at the cutoff times a prognostic contrast. Therefore, an exactly continuous density means no bias, and the population density jump bounds the bias. Valid inference on the jump turns this into a confidence bound under a stated sensitivity assumption. Within this model, bias behind a continuous density requires feedback in both directions. For this case, we bound bias by total boundary movement and construct conservative confidence sets for binary outcomes under explicit sensitivity and smoothness assumptions. We give closed forms for independent uniform scores and for correlated Gaussian scores. In one example, a true risk reduction of four percentage points appears as a one-point increase. Simulations confirm the formulas. Narrower bandwidths, robust bias correction, and machine-learning covariate adjustment do not remove the bias. A fuzzy RD on the first score recovers the causal effect when early and late treatment have the same effect. Our results concern how the running variable is built. We do not assess any published clinical study.
\end{abstract}

\section{Introduction}
Hospitals increasingly use automated risk scores to find patients whose condition may get worse. Early warning scores such as the National Early Warning Score are calculated from vital signs, and an alert is sent when the score crosses a threshold \citep{smith2013,rcp2017}. Electronic health record systems now compute these scores many times during a hospital stay. Large health systems have reported that automated alerts linked to rapid-response teams are associated with lower mortality \citep{escobar2020}. Many of these scores are now produced by machine-learning and other artificial intelligence (AI) models that are updated throughout the hospital stay \citep{escobar2020,gallo2024}. It is important to evaluate these systems carefully, because an alert changes the care a patient receives.

Regression discontinuity (RD) designs are a natural way to evaluate threshold-based alerts. The design was introduced by \citet{thistlethwaite1960} and is now a standard tool for causal inference \citep{imbens2008,leelemieux2010,cattaneo2020book}. Patients just above and just below the threshold should be similar, so the jump in outcomes at the threshold estimates the effect of the alert \citep{lee2008}. Health researchers have argued that RD is underused in medicine, although many clinical decisions follow threshold rules \citep{bor2014,moscoe2015,venkataramani2016}. Recent studies show the value of the design with electronic health records. For example, \citet{eyting2025} used a date-of-birth eligibility rule for herpes zoster vaccination in Wales to estimate the effect of vaccination on dementia.

When a score is measured many times, however, the analyst must choose one running variable. One option is the maximum score during the stay. The maximum shows whether the threshold was ever crossed, so it classifies treatment correctly. But it can also include scores recorded after an earlier alert has already changed care. \citet{gallo2024} evaluated an AI-enabled deterioration alert in this way. They used the maximum score during hospitalization, censored at the outcome, and they noted that the intervention could affect later scores. Later correspondence discussed bandwidth choice and the timing of the first crossing relative to the maximum \citep{goulden2024,gallo2024reply}. We use this design question as motivation only. We do not reanalyze those data or assess the reported clinical effect.

The above discussion implies that the choice of running variable is a design decision with causal consequences. In our present study, we ask what a standard RD identifies when repeated scores are summarized by their maximum and later scores respond to treatment.

\section{Literature Review}
Previous research gives three lines of work that are close to our question.

The first line studies RD designs with several rating scores. \citet{papay2011} extended RD to several assignment variables. When treatment occurs if any score crosses its cutoff, the effect along the treatment frontier is a weighted average of univariate RD effects \citep{wong2013}. One common estimator collapses the scores into a single ``binding score,'' such as the maximum of the centered scores \citep{reardon2012}. The maximum-score analysis in this paper is a binding-score RD. However, this literature assumes that all scores are fixed before assignment. In an alert system, the second score is measured after a first alert may already have changed care.

The second line studies dynamic RD designs, in which units face treatment assignment many times. \citet{cellini2010} developed this design to study school facility investments. \citet{hsushen2024} study identification with repeated treatment opportunities and heterogeneous effects. They define a potential second-period running variable $R_2(d_1)$ that can depend on first-period treatment. In that notation, our feedback is $K=R_2(0)-R_2(1)$. \citet{ruggieri2025} adapts the no-anticipation and common-trends assumptions of difference-in-differences to identify cutoff-specific dynamic effects. Each effect is the sum of two static RD contrasts. Both frameworks define their targets at the cutoff of the first-period running variable. Neither studies what happens when the score sequence is replaced by its maximum. More broadly, our problem is related to adjustment for a variable that is affected by treatment, which is known to bias causal comparisons \citep{rosenbaum1984,montgomery2018}. Feedback between treatment and later measurements is also a central topic in longitudinal causal inference \citep{robins1986,hernan2020}.

The third line studies manipulation of the running variable. \citet{mccrary2008} proposed a test for a jump in the density of the running variable at the cutoff. \citet{cattaneo2020} developed local polynomial density estimators for the same purpose. Researchers now agree that a continuous density is neither necessary nor sufficient for RD validity. \citet{gerard2020} showed that RD effects are only partially identified under one-sided manipulation, and they inferred the extent of manipulation from the density jump at the cutoff. \citet{liao2025} studies running variables distorted by policy. Our one-directional result in Section~\ref{sec:onesign} is close in spirit to \citet{gerard2020}.

\section{Objectives}
That later scores may respond to an alert is not a new idea; \citet{gallo2024} noted it. What is missing is an explicit account of what the maximum-score comparison identifies when this happens. The study has five objectives. First, we write the maximum-score RD contrast as the causal effect plus a difference of two outcome-weighted boundary averages (Proposition~\ref{prop:general}). The formula has the same two-branch structure as the frontier decomposition of \citet{wong2013}, but feedback shifts the treated side of the early-crossing branch. Second, we show that when feedback acts in one direction, the bias equals the relative density jump at the cutoff times a prognostic contrast (Proposition~\ref{prop:onesign}). A continuous density then rules out bias, and the density jump bounds it. Bias behind a continuous density requires feedback of both signs. Third, we give closed forms for independent uniform scores and correlated Gaussian scores (Propositions~\ref{prop:uniform} and~\ref{prop:gauss}). Fourth, we give outer sensitivity bounds for mixed-direction feedback and a conservative finite-sample confidence procedure for binary outcomes (Propositions~\ref{prop:mixed} and~\ref{prop:confidence}). Fifth, we show that a fuzzy RD on the first score recovers the effect when early and late treatment have the same effect. This is a benchmark under stated assumptions, not a general repair. We also use simulations to show that narrower bandwidths, robust bias correction, and flexible covariate adjustment do not remove the bias.

\section{A two-assessment alert model}\label{sec:model}
Let $X$ be the score at the first assessment and $Z$ the score that would be recorded at the second assessment without earlier intervention. Center the threshold at zero. Define
\[
A_0=1\{X\ge 0\},\qquad S_1=Z-KA_0,
\]
where $K$ is the change in the second score caused by an initial intervention. Positive $K$ lowers the recorded score and negative $K$ raises it. Feedback may reflect a true physiological response or a change in measurement, such as more frequent monitoring. It is not assumed to be a direct effect on the outcome.

A patient is treated if an alert occurs at either assessment, so
\[
D=A_0\vee 1\{S_1\ge 0\},\qquad M=\max(X,S_1).
\]
Because the first crossing remains in the treatment history, $D=1\{M\ge0\}=1\{X\ge0 \text{ or } Z\ge0\}$. The maximum therefore classifies treatment correctly. The problem studied below is not misclassification.

Let $Y(d)$ denote potential outcomes and $Y=Y(D)$. The maximum-score RD estimand is
\[
\Delta_M=\lim_{m\downarrow0}E(Y\mid M=m)-\lim_{m\uparrow0}E(Y\mid M=m).
\]
We compare $\Delta_M$ with a constant conditional mean effect $\tau$. Treatment initiated at the first or second assessment has the same effect, and there is no censoring or loss to follow-up. These restrictions let us attribute any gap between $\Delta_M$ and $\tau$ to the construction of the running variable.

\section{Boundary composition under score feedback}
Let $K$ have distribution $G$. Conditional on $K=k$, let $(X,Z)$ have joint density $f_k(x,z)$, and let $g_k(x,z)=E\{Y(0)\mid X=x,Z=z,K=k\}$. Assume $E\{Y(1)-Y(0)\mid X,Z,K\}=\tau$. Assume also continuity at the relevant boundaries, positive finite boundary densities, and integrable envelopes that allow limits to pass under the integrals. For a boundary shift $a(k)$ define
\begin{align*}
\D(a)&=\int\Big\{\int_{-\infty}^{a(k)}f_k(0,z)\,dz+\int_{-\infty}^{0}f_k(x,0)\,dx\Big\}\,dG(k),\\
\N(a)&=\int\Big\{\int_{-\infty}^{a(k)}g_k(0,z)f_k(0,z)\,dz+\int_{-\infty}^{0}g_k(x,0)f_k(x,0)\,dx\Big\}\,dG(k).
\end{align*}
Write $a_0(k)=0$ and $a_K(k)=k$.

\begin{proposition}[Boundary composition]\label{prop:general}
Under the conditions above, $f_M(0-)=\D(a_0)$, $f_M(0+)=\D(a_K)$, and
\[
\Delta_M=\tau+\frac{\N(a_K)}{\D(a_K)}-\frac{\N(a_0)}{\D(a_0)}.
\]
\end{proposition}
Just below zero, either score can be the maximum, and neither assessment has triggered treatment. Just above zero, the initial-score branch admits latent second scores up to $K$ rather than up to zero, because the initial intervention has already shifted them. The late-crossing branch is the same on both sides. The proof is in Appendix~\ref{app:general}.

\begin{remark}
Density continuity makes the two denominators equal. It does not make the numerators equal. A continuous density of the maximum therefore does not establish continuity of untreated outcome means. A sufficient condition for a zero gap is that $g_k$ equals the same constant at every boundary point that enters $\N(a_0)$ and $\N(a_K)$, for $G$-almost every $k$. Independence of $g_k$ from $z$ is not sufficient. Feedback also changes the relative weight of the $X=0$ branch and the $Z=0$ branch, and these branches differ in $X$. For example, with independent uniform scores, $K=0.5$, and $E\{Y(0)\mid X,Z,K\}=0.4+0.2X$, the gap is $0.01$. In the two models of Section~\ref{sec:closed}, the untreated mean depends on $Z$ only, so the gap there vanishes when $\beta=0$.
\end{remark}

\subsection{One-directional feedback}\label{sec:onesign}
Write $f_+=f_M(0+)$, $f_-=f_M(0-)$, and $J=f_+-f_-$. Let $\mu_0=\N(a_0)/\D(a_0)$ be the untreated mean just below the cutoff. By Proposition~\ref{prop:general},
\[
J=\int\Big\{\int_{-\infty}^{k}f_k(0,z)\,dz-\int_{-\infty}^{0}f_k(0,z)\,dz\Big\}\,dG(k),
\]
the boundary mass that feedback adds (if $k>0$) or removes (if $k<0$) on the early-crossing branch.

\begin{proposition}[One-directional feedback]\label{prop:onesign}
Under the conditions of Proposition~\ref{prop:general}, assume $g_k$ is bounded and $K\ge0$ almost surely, or $K\le0$ almost surely. If $J\neq0$, let
\[
m_K=\frac{\N(a_K)-\N(a_0)}{\D(a_K)-\D(a_0)},
\]
the untreated mean over the boundary mass between $0$ and $K$ on the early-crossing branch. Then
\[
\Delta_M-\tau=\frac{J}{f_+}\,(m_K-\mu_0).
\]
If $J=0$, then $\Delta_M=\tau$. If $g_k$ takes values in $[L,U]$ on the boundaries, then $|\Delta_M-\tau|\le(|J|/f_+)(U-L)$. For a binary outcome, $|\Delta_M-\tau|\le|J|/f_+$.
\end{proposition}
The proof is in Appendix~\ref{app:onesign}. Two consequences follow. First, under one-directional feedback, an exactly continuous density of the maximum rules out a gap. A nonzero gap with a continuous density, as in Corollary~\ref{cor:sym}, requires feedback of both signs, with upward and downward shifts in the boundary mass cancelling. This could arise if an alert improves some patients while more intensive monitoring records worse vital signs for others, but we have not shown that this happens in any clinical system. Second, a nonsignificant density test is not the same as $J=0$. A finite-sample test can miss a real jump. The bound uses the size of the jump, not the test decision. The relative jump $j=J/f_+$ can be estimated with local polynomial density methods \citep{cattaneo2020}. Given a valid confidence interval $[\ell,u]$ for $j$, and an assumed bound $|m_K-\mu_0|\le C$, the quantity $B=C\max\{|\ell|,|u|\}$ is a confidence bound for $|\Delta_M-\tau|$. Its coverage holds conditional on the sensitivity assumption and on the validity of the interval for the ratio, including uncertainty in its denominator. Proposition~\ref{prop:onesign} gives the population bound; it does not by itself establish the coverage of any estimation procedure. The contrast $m_K-\mu_0$ compares the untreated prognosis of patients added to or removed from the treated side with that of patients just below the cutoff. It is a sensitivity parameter that analysts can bound from clinical knowledge, or by the outcome range.

\paragraph{Relation to bounds under manipulation.} \citet{gerard2020} derive sharp bounds for RD treatment effects under one-sided manipulation, using the density discontinuity to measure the extent of contamination. Their framework does not require intentional wrongdoing. Proposition~\ref{prop:onesign} uses related logic, but the mathematical structure and target differ: feedback adds or removes mass on one branch of a maximum-score boundary, and the target here is a constant conditional mean effect. We obtain an exact prognostic-contrast identity, rather than sharp bounds from trimming outcome distributions. The implication from density continuity to zero bias depends on this nested boundary structure. It is not a general claim about every monotone distortion of a running variable; \citet{ishihara2024} explain why broader manipulation models need additional restrictions. The contribution below is a sensitivity calculation for opposing boundary movements within the stated feedback model.

\subsection{Mixed-direction sensitivity bounds}\label{sec:mixed}
Define the nonnegative boundary density masses
\[
A_+=\int_{k>0}\int_0^k f_k(0,z)\,dz\,dG(k),\qquad
A_-=\int_{k<0}\int_k^0 f_k(0,z)\,dz\,dG(k).
\]
Let $m_+$ and $m_-$ be the corresponding averages of $g_k(0,z)$, with arbitrary values when the relevant mass is zero. Write $T=A_++A_-$ and $j=J/f_+=1-\lambda$, where $\lambda=f_-/f_+$. Then
\begin{equation}\label{eq:flows}
J=A_+-A_-,\qquad B:=\Delta_M-\tau
=\frac{A_+(m_+-\mu_0)-A_-(m_--\mu_0)}{f_+}.
\end{equation}
The density jump measures the net difference of the movements. It does not identify their sum. In particular, $J=0$ can hold with $T>0$ and $B\ne0$.

\begin{proposition}[Outer sensitivity bound]\label{prop:mixed}
Under Proposition~\ref{prop:general}, suppose all relevant untreated boundary means lie in a specified range $[L,U]$. Let $c=(L+U)/2$ and $W=U-L$. For an assumed upper bound $T/f_+\le\gamma$, necessarily $\gamma\ge|1-\lambda|$, and
\begin{equation}\label{eq:mixedbias}
|B-j(c-\mu_0)|\le\gamma W/2.
\end{equation}
Writing $\mu_-=E(Y\mid M=0-)=\mu_0$ and $\mu_+=E(Y\mid M=0+)$, an outer region for $\tau$ is
\begin{equation}\label{eq:mixedregion}
\mathcal I_\gamma=
\left[\mu_+-c+\lambda(c-\mu_-)-\gamma W/2,\;
      \mu_+-c+\lambda(c-\mu_-)+\gamma W/2\right].
\end{equation}
For binary outcomes, intersect this region with $[-1,1]$.
\end{proposition}
The proof is in Appendix~\ref{app:mixed}. The bound allows arbitrary dependence between feedback and baseline prognosis under the boundary conditions. It does not require the linear outcome model used in Section~\ref{sec:closed}. We do not claim sharpness: compatibility among the shared boundary components and potential-outcome means may tighten the region. The parameter $\gamma$ measures total movement relative to the right-hand boundary density. It is not the proportion of patients affected or the magnitude of their score changes. The observed density supplies the lower constraint $|1-\lambda|\le\gamma$, but cannot supply an upper bound on $\gamma$. Values of $[L,U]$ narrower than the outcome range require external justification.

For the balanced uniform example with $K=\pm1$, $\alpha=0.4$, $\beta=0.2$, and $\tau=-0.04$, the density ratio is one, $\mu_-=0.35$, and $\mu_+=0.36$. The true relative movement is $T/f_+=0.5$. Using $[L,U]=[0.2,0.6]$, the population region at $\gamma=0.5$ is $[-0.09,0.11]$, containing the true effect despite the positive RD contrast. At $\gamma=0.05$ the lower endpoint reaches zero. This is where this bound stops certifying a positive effect; it does not establish the feasibility of every effect in the region. These calculations are most relevant when only $(Y,M)$ are retained. If the initial score is available and the exclusion and timing restrictions hold, Section~\ref{sec:fuzzy} identifies $\tau$ directly.

\subsection{A conservative confidence procedure for binary outcomes}\label{sec:mixedinference}
A population bound does not account for estimation error. We give one finite-sample construction using only independent observations of $(Y,M)$ with $Y\in\{0,1\}$. Choose a fixed bandwidth $h>0$ before examining outcomes. In each window $W_-=(-h,0)$ and $W_+=[0,h)$, let $N_s$ count observations and let $S_s$ count outcomes equal to one. Define $p_s=P(M\in W_s)$ and $q_s=E(Y\mid M\in W_s)$. For a desired error probability $\eta$, form four two-sided Clopper--Pearson intervals \citep{clopper1934}, each with error probability $\eta/4$: two for $p_s$ based on $(N_s,n)$, and two for $q_s$ based on $(S_s,N_s)$. Use $[0,1]$ for a conditional mean with $N_s=0$.

Suppose prespecified constants $H_f,H_\mu$ satisfy, on each side of the cutoff within distance $h$,
\begin{equation}\label{eq:lipschitz}
|f_M(m)-f_s|\le H_f|m|,\qquad
|E(Y\mid M=m)-\mu_s|\le H_\mu|m|.
\end{equation}
For component intervals $[p_s^l,p_s^u]$ and $[q_s^l,q_s^u]$, define
\begin{align}\label{eq:rectangle}
F_s&=\left[\max\{0,p_s^l/h-H_fh/2\},\;p_s^u/h+H_fh/2\right],\\
Q_s&=[q_s^l-H_\mu h,\;q_s^u+H_\mu h]\cap[0,1].\nonumber
\end{align}
Let $\mathcal R=F_-\times F_+\times Q_-\times Q_+$ be the resulting rectangle for $(f_-,f_+,\mu_-,\mu_+)$. For each candidate in this rectangle with positive densities, retain it if $\mu_-\in[L,U]$ and $|1-f_-/f_+|\le\gamma$, and take the union of its regions~\eqref{eq:mixedregion}, intersected with $[-1,1]$. Call this set $\mathcal C_\gamma$. If the lower endpoint of $F_+$ is zero, use $[-1,1]$; if there are no compatible candidates with a strictly positive denominator bound, report the empty set explicitly.

\begin{proposition}[Finite-sample coverage]\label{prop:confidence}
Under the binary-outcome, independence, boundary, and smoothness assumptions above, and the specified range $[L,U]$,
\[
P\{\tau\in\mathcal C_\gamma\}\ge1-\eta
\quad\text{whenever}\quad T/f_+\le\gamma.
\]
The same rectangle covers the entire sensitivity curve simultaneously: with probability at least $1-\eta$, inclusion holds for every $\gamma\ge T/f_+$.
\end{proposition}
Appendix~\ref{app:confidence} proves the result and gives closed-form endpoints. No independence among the four intervals is needed. The guarantee includes smoothing error through~\eqref{eq:lipschitz}; it is conditional on these supplied constants and on the movement and prognostic-range assumptions. Estimating them from the same data without an additional argument does not preserve the guarantee. The construction is deliberately conservative. It is a baseline with transparent coverage, not an efficiency or optimality result, and does not cover dependent patient records or a bandwidth chosen adaptively from outcomes.

\section{Closed forms}\label{sec:closed}
\subsection{Independent uniform scores}
Let $X$ and $Z$ be independent $\mathrm{Uniform}(-1,1)$, let $-1\le K\le 1$ be independent of them, and let $U\sim\mathrm{Uniform}(0,1)$ be independent of all three. Set
\[
Y(d)=1\{U\le\alpha+\beta Z+\tau d\},
\]
with parameters chosen so that all probabilities lie in $[0,1]$. Write $\mu_K=E(K)$ and $\nu_K=E(K^2)$.

\begin{proposition}[Uniform scores]\label{prop:uniform}
In the uniform model, $f_M(0-)=1/2$, $f_M(0+)=(2+\mu_K)/4$, and
\[
\Delta_M=\tau+\beta\,\frac{2\nu_K+\mu_K}{4(2+\mu_K)}.
\]
\end{proposition}
The gap is a property of the estimand, not of estimation. It persists as the sample size grows and the bandwidth shrinks. For common downward feedback $K=k\ge0$, the gap is $\beta k(2k+1)/\{4(2+k)\}$. For common upward feedback $K=-1/4$ it is $-\beta/56$, so the direction is not always toward zero. For $K=-1/2$ the two moments cancel and the gap is zero, although the density jumps. In the notation of Proposition~\ref{prop:onesign}, $m_K=\mu_0$ in that case.

\begin{figure}[htbp]\centering
\includegraphics[width=0.95\linewidth]{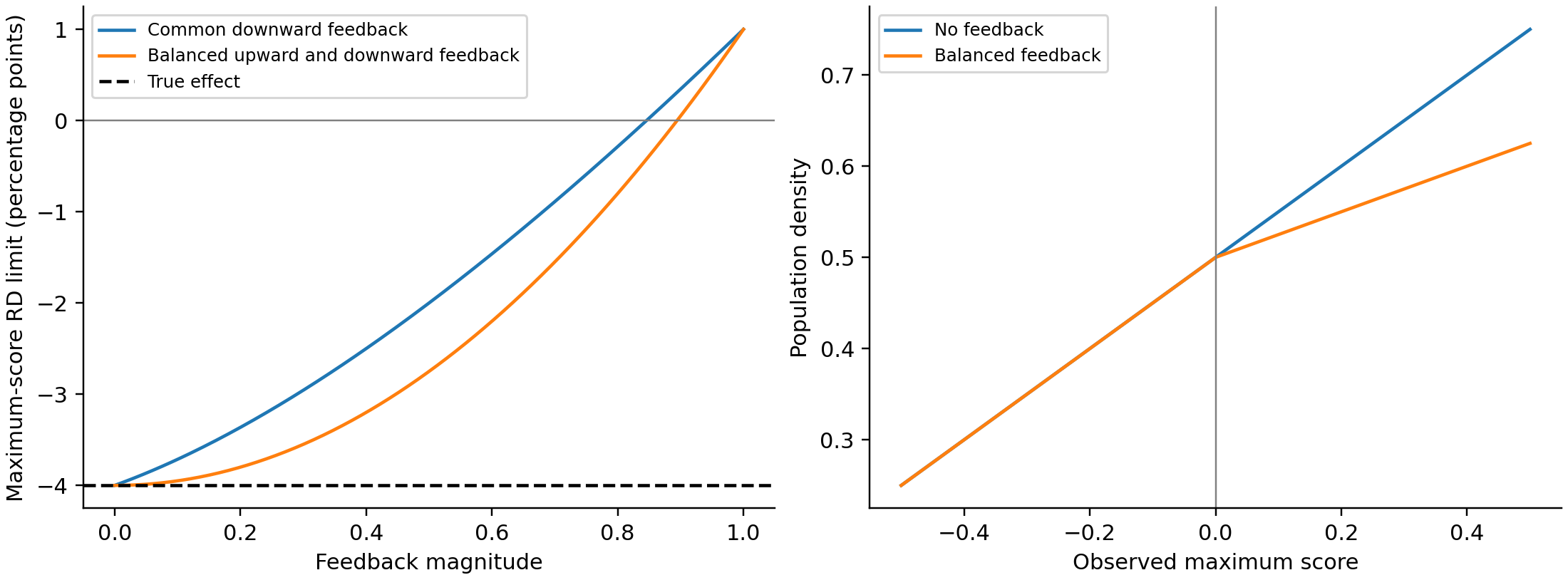}
\caption{Uniform model with $\tau=-0.04$ and $\beta=0.20$. Left: limiting maximum-score RD contrast against feedback magnitude $k$, for common downward feedback $K=k$ and symmetric feedback $K=\pm k$ (Proposition~\ref{prop:uniform}). Right: population density of the maximum with no feedback and with symmetric feedback $K=\pm1$. Both densities are continuous at zero. Symmetric feedback changes the slope of the density above the cutoff, so a test of the density level has nothing to detect, although a test of its derivative might.}\label{fig:pop}
\end{figure}

\begin{corollary}[Symmetric feedback]\label{cor:sym}
If $K=\pm k$ with equal probability, then $f_M(0+)=f_M(0-)$ and $\Delta_M-\tau=\beta k^2/4$.
\end{corollary}
Upward and downward feedback cancel in the density but not in prognostic composition. With $\alpha=0.40$, $\beta=0.20$, $\tau=-0.04$, and $K=\pm1$, the maximum-score contrast is $+0.01$, although the causal risk difference is $-0.04$. With $\tau=0$, the contrast is $0.05$. Figure~\ref{fig:pop} plots the limit against the feedback magnitude and shows the continuous density under symmetric feedback.

\subsection{Correlated Gaussian scores}
The uniform model has independent scores. Real trajectories are correlated, so we check the mechanism under a bivariate normal model. Let $(X,Z)$ be bivariate standard normal with correlation $\rho\in(-1,1)$, independent of $K$. Because $Z$ is unbounded, a linear mean cannot be a probability for all $Z$, so this model uses a continuous outcome,
\[
Y(d)=\alpha+\beta Z+\tau d+\varepsilon,\qquad E(\varepsilon\mid X,Z,K)=0,
\]
measured in arbitrary units. The uniform model keeps its risk-difference interpretation. Write $\sigma=\sqrt{1-\rho^2}$, and let $\phi$ and $\Phi$ be the standard normal density and distribution function.

\begin{proposition}[Correlated Gaussian scores]\label{prop:gauss}
In the Gaussian model,
\[
\Delta_M-\tau=\beta\sigma\left\{\phi(0)-\frac{E\,\phi(K/\sigma)}{\tfrac12+E\,\Phi(K/\sigma)}\right\},\qquad
\frac{f_M(0+)}{f_M(0-)}=\tfrac12+E\,\Phi(K/\sigma).
\]
If $K=\pm k$ with equal probability, the density is continuous at zero and
\[
\Delta_M-\tau=\beta\sigma\{\phi(0)-\phi(k/\sigma)\}>0\quad\text{for }k\neq0,\ \beta>0.
\]
\end{proposition}
The proof is in Appendix~\ref{app:gauss}. The gap does not depend on the uniform assumption. It vanishes as $|\rho|\to1$, because the latent second score of a patient at $X=0$ then concentrates near zero, so feedback can shift little composition. The gap is not monotone in $\rho$; for $k=1$ it peaks near $\rho=0.8$. For moderate correlation it is close to the independent case. With $\beta=0.20$ and $k=1$, the gap is 0.031, 0.034, and 0.036 outcome units for $\rho=0$, $0.5$, and $0.7$.

We checked the formula in two ways. First, we integrated the boundary expressions of Proposition~\ref{prop:general} numerically; they agreed with the closed form to at least five digits. Second, we ran a Monte Carlo study with $\alpha=0.40$, $\beta=0.20$, $\tau=-0.04$, $\varepsilon\sim N(0,0.3^2)$, $n=10^6$, 200 replications per setting, and seed 20261115. Each replication fit separate triangular-kernel local linear regressions on each side of $M=0$ with bandwidth $0.025$. Table~\ref{tab:gauss} reports the results. Every Monte Carlo mean was within 1.8 Monte Carlo standard errors of the analytic gap. The ratio of empirical densities in $[0,0.005)$ and $[-0.005,0)$ was within about 0.5\% of one in every setting.

\begin{table}[htbp]\centering\small
\caption{Gaussian check: gap $\Delta_M-\tau$ in outcome units (200 replications, $n=10^6$, $h=0.025$).}\label{tab:gauss}
\begin{tabular}{rrrrr}\toprule
$\rho$ & $k$ & Analytic gap & Monte Carlo mean & MC SE\\\midrule
0.0 & 1.0 & 0.0314 & 0.0315 & 0.0007\\
0.5 & 1.0 & 0.0336 & 0.0339 & 0.0007\\
0.8 & 1.0 & 0.0359 & 0.0347 & 0.0007\\
0.5 & 0.5 & 0.0106 & 0.0107 & 0.0007\\\bottomrule
\end{tabular}
\end{table}

\section{A benchmark using the initial score}\label{sec:fuzzy}
The initial score is measured before any intervention.

\begin{proposition}[Initial-score fuzzy RD]\label{prop:fuzzy}
In either model above, with outcomes as defined there, if $f_k(x,z)$ is continuous in $x$ at $x=0$ for each $z$ and $k$, the untreated outcome mean is continuous at $X=0$. The first-stage jump is $P(D=1\mid X=0+)-P(D=1\mid X=0-)=P(Z<0\mid X=0)$. The Wald ratio of the outcome jump to the treatment jump equals $\tau$.
\end{proposition}
In the uniform model the first-stage jump is $1/2$ and the outcome jump is $\tau/2$. The sharp initial-score contrast estimates the effect of initial eligibility when patients below the cutoff can still be treated later. Dividing by the first-stage jump, as in a standard fuzzy RD \citep{hahn2001}, makes it comparable with the treatment effect. This requires the exclusion and timing restrictions of the model. The initial-score analysis follows the same logic as the dynamic RD frameworks above, which condition on the first-period running variable at its cutoff. If early treatment has a different effect from late treatment, the initial-score contrast mixes a timing effect with the treatment effect. The history-specific estimands of \citet{hsushen2024} and \citet{ruggieri2025} are then the appropriate targets.

\section{Simulation study}\label{sec:sim}
We generated 1,000 samples of $n=20{,}000$ independent individuals for each of nine scenarios of the uniform model, with $\alpha=0.40$ and varying feedback, $\beta$, and $\tau$. Common random numbers paired the scenarios, so initial-score estimates are identical across scenarios with the same outcome parameters. We fit separate local linear regressions on each side with triangular kernels and prespecified bandwidths of 0.10 and 0.20. For the fuzzy design we estimated the outcome and uptake jumps jointly and used a delta-method standard error. Intervals use HC1 standard errors.

We also integrated the population weighted least-squares equations numerically. This gives the exact finite-bandwidth target for each estimator. Sampling error, finite-bandwidth approximation error, and the causal gap can then be separated. The boundary formulas were checked independently at distances down to $10^{-8}$ from the cutoff. The largest discrepancy was $9.2\times10^{-9}$.

\begin{table}[t]\centering\small
\caption{Simulation results at bandwidth 0.20 (percentage points; 1,000 replications, $n=20{,}000$).}\label{tab:sim}
\begin{tabular}{lrrrrrr}\toprule
& & \multicolumn{3}{c}{Maximum-score sharp RD} & \multicolumn{2}{c}{Initial-score fuzzy RD}\\\cmidrule(lr){3-5}\cmidrule(lr){6-7}
Scenario & $\tau$ & Limit & Mean & Cover. & Mean & Cover.\\\midrule
No feedback & -4.00 & -4.00 & -3.91 & 95.4 & -3.91 & 95.8 \\
Downward 0.5 & -4.00 & -2.00 & -1.88 & 89.9 & -3.91 & 95.8 \\
Downward 1.0 & -4.00 & 1.00 & 1.13 & 62.3 & -3.91 & 95.8 \\
Upward 0.25 & -4.00 & -4.36 & -4.32 & 95.4 & -3.91 & 95.8 \\
Upward 0.5 & -4.00 & -4.00 & -4.01 & 94.9 & -3.91 & 95.8 \\
Balanced $\pm$ 0.5 & -4.00 & -2.75 & -2.71 & 92.8 & -3.91 & 95.8 \\
Balanced $\pm$ 1.0 & -4.00 & 1.00 & 1.05 & 68.8 & -3.91 & 95.8 \\
Balanced null outcome effect & 0.00 & 5.00 & 5.04 & 69.3 & 0.01 & 95.4 \\
Balanced nonprognostic marker & -4.00 & -4.00 & -3.95 & 94.6 & -3.86 & 95.4 \\\bottomrule\end{tabular}
\par\smallskip\parbox{0.95\linewidth}{\footnotesize Limit is the population RD contrast from Proposition~\ref{prop:uniform}. Coverage is the share of nominal 95\% intervals containing $\tau$. Downward and upward rows use a common $K$; balanced rows use $K=\pm k$. The nonprognostic row sets $\beta=0$ with $K=\pm1$.}
\end{table}

\begin{figure}[htbp]\centering
\includegraphics[width=0.95\linewidth]{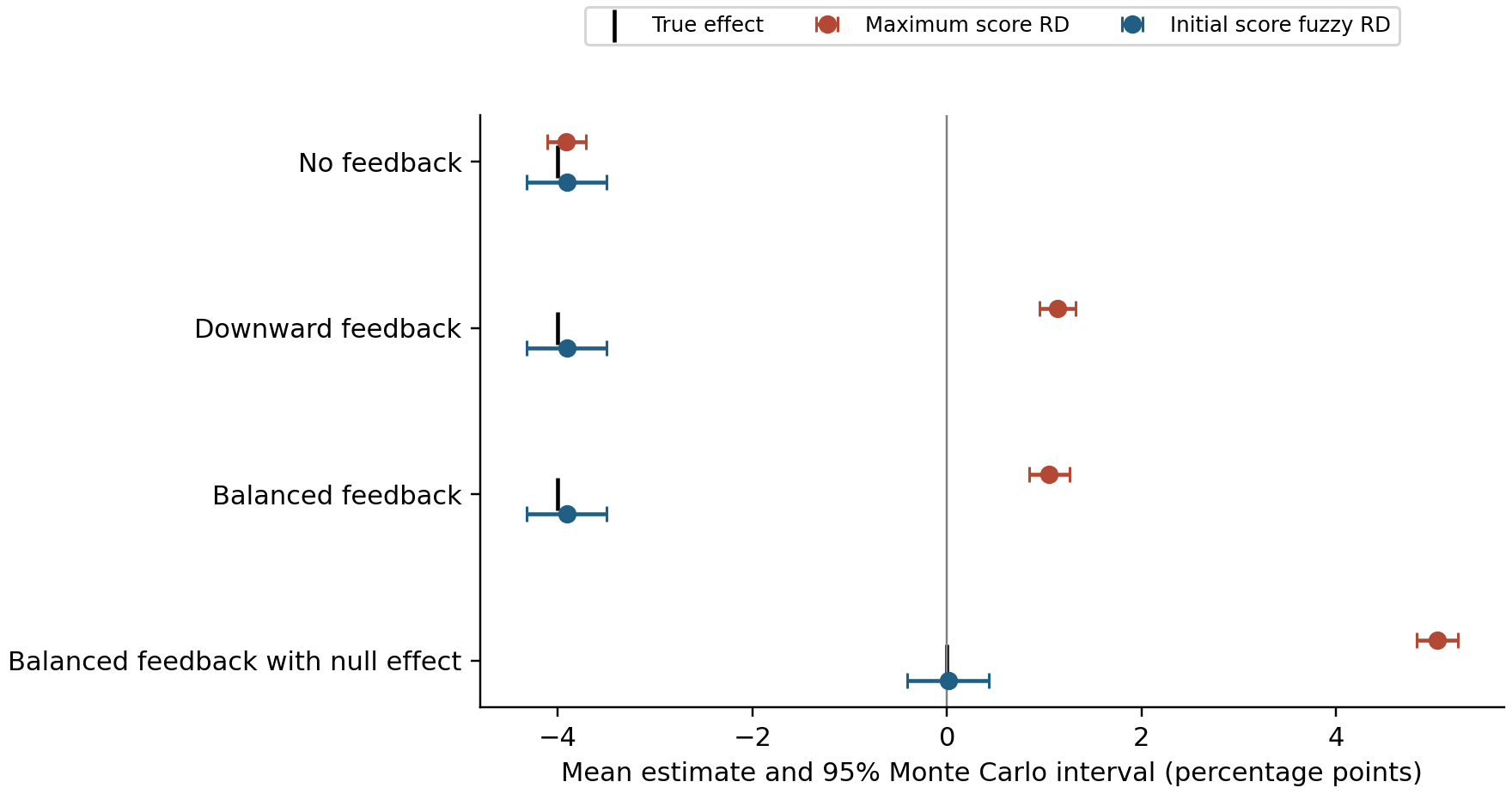}
\caption{Monte Carlo means at bandwidth 0.20 with $\pm1.96$ Monte Carlo standard errors (1,000 replications, $n=20{,}000$). Downward feedback is $K=1$; balanced feedback is $K=\pm1$. Black marks show the causal risk differences. The intervals describe uncertainty in the simulation means, not in a single study's estimate.}\label{fig:sim}
\end{figure}

Table~\ref{tab:sim} reports results at bandwidth 0.20, and Figure~\ref{fig:sim} plots selected scenarios. Monte Carlo means match the analytic limits in every scenario. With $K=\pm1$ and $\tau=-0.04$, the mean maximum-score estimate was $+1.05$ points (Monte Carlo SE 0.11). The initial-score fuzzy estimate averaged $-3.91$ points (Monte Carlo SE 0.21). In the null scenario, the maximum-score analysis rejected a zero effect in 30.7\% of replications at nominal 5\%. The fuzzy analysis rejected in 4.6\%. Maximum-score intervals covered their own finite-bandwidth target in 94.4\% of replications, but covered the causal null in only 69.3\%. The intervals are accurate for the regression target; the target is the wrong one.

At bandwidth 0.10, rejection in the null scenario fell to 18.0\% while the mean stayed near five points. A smaller bandwidth increased variance; it did not remove the gap. Across all scenarios, methods, and bandwidths, each Monte Carlo mean was within 1.42 Monte Carlo standard errors of its population finite-bandwidth target.

\paragraph{Bias correction and flexible covariate adjustment.} We also ran 100 replications with $n=6{,}000$ and a continuous outcome in arbitrary units, with $\tau=-0.04$. We used \texttt{rdrobust} \citep{calonico2014} with robust bias-corrected inference and the RDFlex covariate adjustment of \citet{noack2026} with gradient-boosted nuisance models. Appendix~\ref{app:ml} gives the full specification. Table~\ref{tab:ml} reports the results.

\begin{table}[htbp]\centering\small
\caption{Robust bias-corrected and RDFlex estimates (continuous outcome, $\tau=-0.04$; 100 replications, $n=6{,}000$).}\label{tab:ml}
\begin{tabular}{llrrr}\toprule
Feedback & Estimator & Mean & Mean SE & Coverage (\%)\\\midrule
None & Maximum score, \texttt{rdrobust} & $-0.041$ & 0.070 & 97\\
None & Maximum score, RDFlex & $-0.042$ & 0.017 & 89\\
None & Initial-score fuzzy, \texttt{rdrobust} & $-0.036$ & 0.147 & 97\\
None & Initial-score fuzzy, RDFlex & $-0.047$ & 0.047 & 96\\
$K=\pm1$ & Maximum score, \texttt{rdrobust} & $0.007$ & 0.071 & 94\\
$K=\pm1$ & Maximum score, RDFlex & $0.008$ & 0.018 & 30\\\bottomrule
\end{tabular}
\par\smallskip\parbox{0.9\linewidth}{\footnotesize Initial-score estimates do not depend on feedback and are shown once. Coverage is the share of nominal 95\% intervals containing $\tau$; its Monte Carlo SE is about 2--3 percentage points (about 4.6 for the 30\% entry).}
\end{table}

Under feedback, both maximum-score estimators centered near the RD limit of $0.01$ rather than near $\tau$. The \texttt{rdrobust} intervals still covered $\tau$ in 94\% of replications, but only because they were wide relative to the gap. RDFlex reduced the standard error by about a factor of four, and coverage fell to 30\%. RDFlex coverage was also below nominal without feedback (89\%), so part of the shortfall reflects interval calibration at this sample size. Coverage fell from 89\% without feedback to 30\% with it. The covariates here improve precision only; they carry no information about the selection mechanism. Bias correction targets polynomial approximation error, and covariate adjustment targets variance. Neither restores the missing identification condition. With 100 replications, these results are implementation checks, not a coverage study.

\subsection{Mixed-direction sensitivity and inference}\label{sec:mixedsim}
We evaluated the binary-outcome procedure of Section~\ref{sec:mixedinference} under four uniform-model scenarios: no feedback; $K=0.5$; balanced feedback $P(K=-1)=P(K=1)=0.5$; and nearly balanced feedback $P(K=-1)=0.45$, $P(K=1)=0.55$. All use $\alpha=0.4$, $\beta=0.2$, and $\tau=-0.04$. There were 1,000 replications at each of $n=20{,}000$ and $n=200{,}000$, with fixed bandwidths $h=0.10$ and $h=0.05$, respectively, and seed 2026111501. Common random numbers paired scenarios within each replication. The confidence procedure used only $(Y,M)$; latent quantities were used only to validate assumptions and evaluate performance.

We supplied $[L,U]=[0.2,0.6]$, $H_f=0.5$, and $H_\mu=0.25$. The prognostic range holds by construction. Appendix~\ref{app:mixedsim} gives the exact within-window density and mean functions verifying the smoothness bounds. The true movement ratios are $0$, $0.20$, $0.50$, and $10/21\simeq0.476$, respectively. We evaluated $\gamma\in\{0,0.10,0.20,0.50,0.75\}$ and also the unrestricted binary prognostic range $[0,1]$. The ancillary files retain every replication, including empty sets as noncoverage; no failed set is omitted.

\begin{table}[htbp]\centering\small
\caption{Sensitivity confidence sets and conventional maximum-score intervals (1,000 replications per row). Width is in percentage points; coverage and wrong-sign rates are percentages.}\label{tab:mixed}
\setlength{\tabcolsep}{4pt}
\begin{tabular}{lrrrrrr}\toprule
& & & \multicolumn{2}{c}{Sensitivity set} & \multicolumn{2}{c}{Conventional RD}\\
\cmidrule(lr){4-5}\cmidrule(lr){6-7}
Scenario & $n$ & $\gamma$ & Cover. & Width & Cover. & Wrong sign\\\midrule
No feedback & 20,000 & 0.00 & 100.0 & 25.0 & 96.1 & 0.2 \\
Downward $0.5$ & 20,000 & 0.20 & 100.0 & 31.5 & 91.4 & 0.9 \\
Balanced $\pm1$ & 20,000 & 0.50 & 100.0 & 48.1 & 81.5 & 4.5 \\
Nearly balanced $\pm1$ & 20,000 & 0.50 & 100.0 & 47.1 & 80.9 & 4.4 \\
No feedback & 200,000 & 0.00 & 100.0 & 11.7 & 94.4 & 0.0 \\
Downward $0.5$ & 200,000 & 0.20 & 100.0 & 18.7 & 81.7 & 0.2 \\
Balanced $\pm1$ & 200,000 & 0.50 & 100.0 & 32.9 & 32.5 & 7.0 \\
Nearly balanced $\pm1$ & 200,000 & 0.50 & 100.0 & 32.5 & 31.9 & 6.6 \\
\bottomrule\end{tabular}
\par\smallskip\parbox{0.97\linewidth}{\footnotesize All selected $\gamma$ values bound the true movement ratio. Nominal coverage is 95\%. Conventional intervals use triangular-kernel local linear fits, HC1 standard errors, and the same $h$ as the sensitivity procedure. ``Wrong sign'' means a lower endpoint above zero despite $\tau=-0.04$. The sensitivity sets had no empty, full-range, or wrong-sign sets in these rows.}
\end{table}

Table~\ref{tab:mixed} shows coverage of 100\% for the sensitivity sets in all eight displayed settings. This is evidence of substantial conservatism, not evidence that coverage is calibrated at 95\%. With $n=200{,}000$ and balanced feedback, the mean set width is 32.9 percentage points, compared with the population outer-region width of 20 points. Conventional intervals cover the causal effect in only 32.5\% of replications and report the wrong sign in 7.0\%. Their Monte Carlo standard errors are 1.48 and 0.81 percentage points, respectively. At $n=20{,}000$, the corresponding rates are 81.5\% and 4.5\%. Greater precision makes the gap between the RD contrast and the causal target more consequential.

\begin{figure}[htbp]\centering
\includegraphics[width=\linewidth]{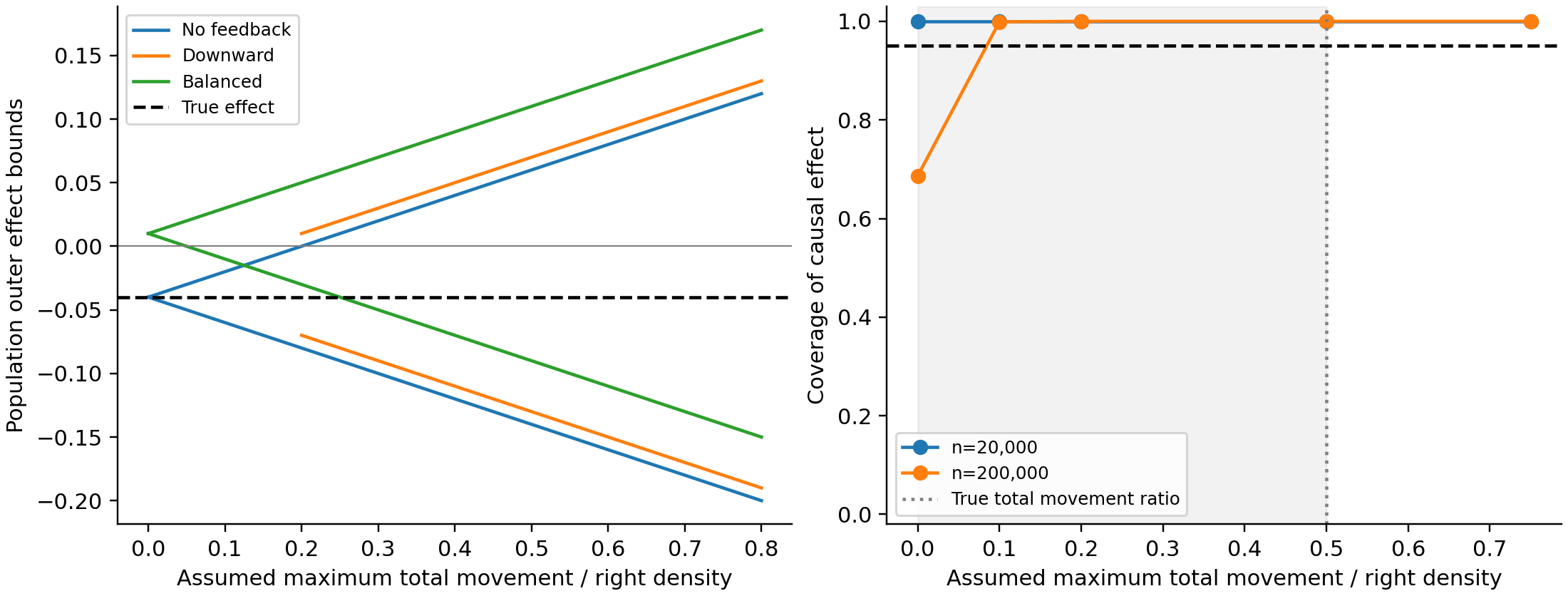}
\caption{Mixed-direction sensitivity analysis with $[L,U]=[0.2,0.6]$. Left: population outer-bound endpoints against $\gamma$, drawn where the observed net-density constraint is satisfied. The true movement ratios are 0 (none), 0.2 (downward), and 0.5 (balanced); assumptions below these values are false in the displayed data-generating processes. The curves need not be sharp compatible regions. Right: empirical coverage for balanced feedback (1,000 replications per sample size); the shaded region uses an underspecified movement bound. Coverage is guaranteed only when $\gamma\ge0.5$ here. The dashed horizontal line is 95\%.}\label{fig:mixed}
\end{figure}

The sensitivity assumptions matter. In the balanced $n=200{,}000$ scenario, incorrectly setting $\gamma=0$ reduces coverage to 68.6\%; the theorem does not apply there. Some other underspecified values still cover because the procedure is conservative, which does not validate those assumptions. Using only $[L,U]=[0,1]$ at the valid $\gamma=0.5$ increases the mean width to 65.2 points. Figure~\ref{fig:mixed} displays the population sensitivity curves and the coverage curve. These are synthetic validations of the mathematics and implementation. They do not establish clinically plausible sensitivity values or favorable efficiency.

\section{Discussion}
Our results raise a specific design question for alert evaluations. Does the running variable summarize a score trajectory that can respond to the intervention being evaluated? Correct classification of treatment does not answer this question. A density test does not answer it either. Bandwidth sensitivity and feedback sensitivity are different problems. A narrower bandwidth reduces approximation error, but it cannot remove a composition difference in the limiting comparison.

We suggest two practical steps, each with limits. First, when the score sequence is available, analysts can report an RD on a pre-intervention score, such as the first measurement or the score at the first possible alert, next to any maximum-score analysis. Disagreement beyond sampling error is consistent with feedback. However, it does not diagnose feedback uniquely. A separate timing effect, or different complier populations under heterogeneous effects, can also produce it. Agreement does not rule out feedback either, because the bias can be small for some feedback distributions.

Second, Proposition~\ref{prop:onesign} gives a calculation that analysts can carry out when feedback plausibly acts in one direction, for example when an alert can only lower later scores. Obtain a valid confidence interval $[\ell,u]$ for the relative density jump $J/f_+$ of the maximum at the cutoff, and choose a defensible bound $C$ on $|m_K-\mu_0|$. Then $C\max\{|\ell|,|u|\}$ is a confidence bound for the bias of the maximum-score contrast, conditional on that assumption. The absolute value matters. Under upward feedback the interval lies below zero, and its upper endpoint does not bound the magnitude. In the downward-feedback scenario with $K=0.5$ in Table~\ref{tab:sim}, $J/f_+=0.2$ and $m_K-\mu_0=0.10$, so the bias is $0.02$. This is the difference between the RD limit ($-2.00$ points) and $\tau$ ($-4.00$ points). An analyst willing to assume $|m_K-\mu_0|\le0.10$ would bound the bias at two percentage points. The outcome-range bound alone gives a much looser 20 points.

When feedback can take both signs, the density jump alone does not bound the bias. Proposition~\ref{prop:mixed} instead asks the analyst to specify an upper bound on total relative boundary movement and a range for untreated prognosis. A useful report would show how the causal region and its confidence set change over these assumptions, alongside the net density jump. The binary-outcome procedure accounts jointly for outcome and density uncertainty. Its wide intervals in the simulations are a substantive result: a large sample cannot remove uncertainty about total movement that the observed maximum does not identify. A defensible application needs external information about score response, clinical prognosis, and local smoothness, or a justified initial-score design. The bounds do not make the missing information available from the maximum alone.

Our results have a specific implication for AI-based alert scores. Machine-learning risk models take their inputs from the electronic health record, and these records reflect both the patient's health and the patient's contact with the healthcare system. \citet{agniel2018} showed that the presence and timing of laboratory test orders, apart from the test results, predict survival. After an alert, clinicians usually order more tests, measure vital signs more often, and start new treatments. A model that reads these records will produce later scores that respond to the alert itself. The response can go in both directions. Successful treatment can lower the later score, while more frequent measurement and new orders can raise it. This is the two-signed feedback of Corollary~\ref{cor:sym} and Section~\ref{sec:mixed}, the case in which a density check is uninformative and the sensitivity bounds of Proposition~\ref{prop:mixed} are needed. Therefore, evaluations of AI-based alerts should be especially careful with running variables built from post-alert scores. Flexible machine-learning adjustment in the estimation step does not solve this problem, as Section~\ref{sec:sim} shows.

We acknowledge several limitations of this study. The model is simple. It has two assessments, no noncompliance, and no timing effect. Symmetric feedback is a constructed mechanism, and we have not shown that it occurs in any clinical system. Event-dependent censoring, discharge, irregular measurement, and alert-induced monitoring are absent, and each may add further selection. Under heterogeneous effects, the maximum-score and initial-score contrasts can also weight different populations, so the target should be defined before the score summary is chosen. The outer bound need not be sharp, and the confidence procedure may be substantially conservative. Future research should study sharper compatible bounds, less conservative inference, longer trajectories with first-crossing designs, explicit early and late treatment effects, and an application with score and alert timestamps.

\section{Conclusion}
A maximum score classifies alert exposure correctly. However, when later scores respond to earlier alerts, the maximum-score RD contrast identifies the causal effect only if the untreated outcome means on the two boundary compositions agree. Feedback together with a prognostic later score makes this condition fail in general, although cancellation can occur, as in the uniform model with $K=-1/2$. When the condition fails, the contrast differs from the causal effect by a composition term. This term does not shrink with sample size or bandwidth, and it can exist together with a continuous running-variable density. We recommend that analysts who evaluate threshold-based alerts with repeated scores state which score defines the comparison, report a pre-intervention-score analysis where the data allow, and treat a maximum-score estimate as causal only with an argument that the untreated outcome means under the two boundary compositions agree. Absent or nonprognostic feedback is sufficient for this, but not necessary. When feedback is one-directional, the density jump at the cutoff and Proposition~\ref{prop:onesign} bound the possible bias. For mixed-direction feedback, the total-movement sensitivity region makes the additional assumptions explicit, and conservative inference is available for independent binary outcomes. The initial-score benchmark is valid under the timing restriction stated here. With timing effects, dynamic RD estimands are the appropriate targets. We hope these findings help researchers and clinicians design more careful evaluations of clinical alert systems.

\paragraph{Funding and competing interests.} The author received no specific funding for this work and declares no competing interests.

\paragraph{Reproducibility.} Ancillary scripts reproduce all tables and figures. File names, seeds, package versions, execution commands, and replication-level sensitivity results are documented in \texttt{anc/README.txt}.

\clearpage
\bibliographystyle{apalike}

\clearpage
\appendix
\section{Proofs}
\subsection{Proposition~\ref{prop:general}}\label{app:general}
Conditional on $K=k$, for $m<0$,
\[P(M\le m\mid k)=\int_{-\infty}^{m}\!\int_{-\infty}^{m}f_k(x,z)\,dz\,dx,\]
and for $m>0$,
\[P(M\le m\mid k)=\int_{-\infty}^{0}\!\int_{-\infty}^{m}f_k(x,z)\,dz\,dx+\int_{0}^{m}\!\int_{-\infty}^{m+k}f_k(x,z)\,dz\,dx.\]
Differentiating, the left density has two parts: the boundary $X=m$ with $Z\le m$, and the boundary $Z=m$ with $X\le m$. On the right, the first integral gives the late-crossing boundary $Z=m$, $X<0$. The second gives the early-crossing boundary $X=m$, $Z\le m+k$. The remaining term over $0<X<m$ vanishes as $m\downarrow0$ under the domination conditions. Taking limits and integrating over $G$ gives $\D(a_0)$ and $\D(a_K)$. Replacing $f_k$ by $g_kf_k$ gives the untreated outcome moments $\N(a_0)$ and $\N(a_K)$. Since $D=0$ below and $D=1$ above the cutoff, the constant effect adds $\tau$ to the right-side mean only.

\subsection{Proposition~\ref{prop:onesign}}\label{app:onesign}
Write $\D_K=\D(a_K)$, $\D_0=\D(a_0)$, and similarly for $\N$. By Proposition~\ref{prop:general},
\[
\Delta_M-\tau=\frac{\N_K}{\D_K}-\frac{\N_0}{\D_0}=\frac{(\N_K-\N_0)-\mu_0(\D_K-\D_0)}{\D_K}.
\]
Here $\D_K-\D_0=J$ and $\D_K=f_+$. This identity holds for any $G$. If $K\ge0$ almost surely, $\N_K-\N_0=\int\int_0^k g_k(0,z)f_k(0,z)\,dz\,dG(k)$ and $J=\int\int_0^k f_k(0,z)\,dz\,dG(k)\ge0$. If $J>0$, $\N_K-\N_0=m_KJ$ with $m_K$ a weighted average of $g_k$ over these points, which gives the formula. If $J=0$, the weight $f_k(0,z)\,dz\,dG(k)$ on $\{0<z<k\}$ is zero, so $\N_K-\N_0=0$ because $g_k$ is bounded, and $\Delta_M=\tau$. The case $K\le0$ is the same with the integrals over $\{k<z<0\}$ and both differences negative. Since $m_K$ and $\mu_0$ are weighted averages of $g_k$ on the boundaries, $|m_K-\mu_0|\le U-L$.

\subsection{Proposition~\ref{prop:mixed}}\label{app:mixed}
Splitting the signed integral in Appendix~\ref{app:onesign} at $k=0$ gives~\eqref{eq:flows}. Subtracting $j(c-\mu_0)$ from $B$ gives
\[
B-j(c-\mu_0)=\frac{A_+(m_+-c)-A_-(m_--c)}{f_+}.
\]
Both $m_+$ and $m_-$ lie in $[L,U]$, so its absolute value is at most $TW/(2f_+)\le\gamma W/2$. Also $|J|\le T$. Since $\tau=\mu_+-\mu_--B$ and $j=1-\lambda$, rearranging gives~\eqref{eq:mixedregion}. The proof supplies an outer bound only; it does not construct a full compatible latent distribution attaining each endpoint.

\subsection{Proposition~\ref{prop:confidence}}\label{app:confidence}
Each $N_s$ is binomial with parameters $(n,p_s)$, and $S_s$ given $N_s$ is binomial with parameters $(N_s,q_s)$. Exact conditional coverage implies unconditional coverage for each conditional-mean interval, including the convention at zero counts. A union bound gives joint coverage of the four window quantities with probability at least $1-\eta$. No joint independence is required. Integrating the density inequality over a window yields $|p_s/h-f_s|\le H_fh/2$. The conditional mean is a density-weighted average in that window, so $|q_s-\mu_s|\le H_\mu h$. Thus the same event places all four cutoff quantities in $\mathcal R$. On that event the true tuple is compatible whenever the sensitivity assumptions hold. Proposition~\ref{prop:mixed} then places $\tau$ in its projected region, for every admissible $\gamma$ on the same event. Returning $[-1,1]$ when the denominator is not bounded away from zero preserves coverage. An empty set is retained as such and counts as noncoverage; under valid assumptions it can occur only off the joint-coverage event.

For implementation, write the endpoints of $F_s$ as $f_s^l,f_s^u$ and of $Q_s$ as $q_s^l,q_s^u$. If $f_+^l>0$, set
\[
\lambda_l=\max\{f_-^l/f_+^u,0,1-\gamma\},\qquad
\lambda_u=\min\{f_-^u/f_+^l,1+\gamma\},
\]
and $a=\max(q_-^l,L)$, $b=\min(q_-^u,U)$. If $\lambda_l>\lambda_u$ or $a>b$, the set is empty. Otherwise its endpoints before intersection with $[-1,1]$ are
\begin{align*}
\ell_\gamma&=q_+^l-c+\min\{\lambda_l(c-b),\lambda_u(c-b)\}-\gamma W/2,\\
u_\gamma&=q_+^u-c+\max\{\lambda_l(c-a),\lambda_u(c-a)\}+\gamma W/2.
\end{align*}
These formulas project the rectangle into the outer region; they are not a sharp latent-model identified set.

\subsection{Proposition~\ref{prop:uniform}}
For fixed $k$, each left boundary contributes density $1/4$, so $f_M(0-)=1/2$. The $X=0$ branch contributes $Z$-moment $\int_{-1}^{0}z/4\,dz=-1/8$, and the $Z=0$ branch contributes zero, so $E(Z\mid M=0-)=-1/4$. On the right, the $X=0$ branch contributes density $(1+k)/4$ and moment $(k^2-1)/8$, and the late-crossing branch contributes $1/4$ and zero. Averaging over $K$ gives $f_M(0+)=(2+\mu_K)/4$ and $E(Z\mid M=0+)=(\nu_K-1)/\{2(2+\mu_K)\}$. Then
\[E(Z\mid M=0+)-E(Z\mid M=0-)=\frac{2\nu_K+\mu_K}{4(2+\mu_K)},\]
and multiplying by $\beta$ gives the result. For $K=\pm k$, $\mu_K=0$ and $\nu_K=k^2$.

\subsection{Proposition~\ref{prop:gauss}}\label{app:gauss}
Given $X=0$, $Z\sim N(0,\sigma^2)$, so $f(0,z)=\phi(0)\phi(z/\sigma)/\sigma$. Hence $\int_{-\infty}^{a}f(0,z)\,dz=\phi(0)\Phi(a/\sigma)$ and $\int_{-\infty}^{a}zf(0,z)\,dz=-\phi(0)\sigma\phi(a/\sigma)$. The $Z=0$ branch has density $\phi(0)/2$ and zero $Z$-moment. Thus $\D(a_0)=\phi(0)$, $\D(a_K)=\phi(0)\{1/2+E\Phi(K/\sigma)\}$, $E(Z\mid M=0-)=-\sigma\phi(0)$, and $E(Z\mid M=0+)=-\sigma E\phi(K/\sigma)/\{1/2+E\Phi(K/\sigma)\}$. Proposition~\ref{prop:general} with $g=\alpha+\beta z$ gives the result. For $K=\pm k$, $E\Phi(K/\sigma)=1/2$ by symmetry and $E\phi(K/\sigma)=\phi(k/\sigma)<\phi(0)$.

\subsection{Proposition~\ref{prop:fuzzy}}
If $X<0$, treatment occurs exactly when $Z\ge0$. If $X\ge0$, treatment has already occurred. The conditional distribution of $(Z,K)$ given $X=x$ is continuous at $x=0$, so the untreated outcome mean is continuous there. The treatment jump is $P(Z<0\mid X=0)$, and the outcome jump is $\tau P(Z<0\mid X=0)$ because the effect does not depend on timing. The ratio is $\tau$.

\section{Specification of the RDFlex and rdrobust experiment}\label{app:ml}
Scores $X,Z\sim\mathrm{Uniform}(-1,1)$ independently, with six covariates $V_1,\dots,V_6\sim\mathrm{Uniform}(-1,1)$ independent of the scores. Treatment is $D=1\{X\ge0 \text{ or } Z\ge0\}$. The outcome is
\[
Y=0.4+0.2Z-0.04D+b(V)+\varepsilon,\quad b(V)=0.6\sin(\pi V_1)+0.4V_2V_3+0.3(V_4^2-1/3),\quad \varepsilon\sim N(0,0.05^2).
\]
$V_5$ and $V_6$ are pure noise. The covariates are prognostic only, so adjustment can reduce variance but cannot address feedback. Feedback is $K=\pm1$ with equal probability, independent of everything else. The no-feedback scenario uses $K=0$ on the same draws.

Conventional estimates use \texttt{rdrobust} 2.1.1 with $h=0.20$, $b=0.30$, $p=1$, $q=2$, a triangular kernel, and HC1 variance. We report the bias-corrected point estimate with the robust standard error and interval. RDFlex uses DoubleML 0.11.4 with three folds, one sample split, one iteration, a first-stage bandwidth of 0.50, and the ``interacted cutoff and score'' first-stage specification. Its final step uses the same \texttt{rdrobust} settings. Outcome nuisances use scikit-learn 1.8.0 \texttt{HistGradientBoostingRegressor} (120 iterations, 15 leaves, learning rate 0.08, minimum leaf size 30, $L_2$ penalty 1). The uptake nuisance in the fuzzy design uses \texttt{HistGradientBoostingClassifier} (100 iterations, 7 leaves, learning rate 0.06, minimum leaf size 40, $L_2$ penalty 2). Early stopping is off. There are 100 replications with $n=6{,}000$ and seed 20261114. The same initial-score fits serve both feedback scenarios.

\section{Smoothness verification for the binary sensitivity experiment}\label{app:mixedsim}
On the left window, $f_M(m)=(1+m)/2$ and $E(Z\mid M=m)=(3m-1)/4$, so the density and observed-mean slopes have magnitudes $0.5$ and $0.15$. For $0\le m\le0.1$ and fixed $k\in\{-1,0,0.5\}$, the right density and unnormalized $Z$ moment are
\[
f_k^+(m)=(2+k)/4+m/2,\qquad
z_k^+(m)=(k^2-1)/8+(k/2+1/4)m+3m^2/8.
\]
For $k=1$, the upper support is saturated, giving $f_1^+(m)=3/4$ and $z_1^+(m)=m/4$. Mixtures average these density and moment polynomials with the specified probabilities. The right observed mean is $0.4-0.04+0.2\,z^+(m)/f^+(m)$. Evaluating the derivative at the endpoints and at any real roots of its derivative within the window gives maximum absolute mean slopes of $0.15$ (none), $0.184$ (downward), and at most $0.15$ for either two-point mixture, including the left side. Thus the supplied $H_f=0.5$ and $H_\mu=0.25$ are valid on both windows. These are analytic checks for the specified data-generating process, not estimates intended for a real application. The script independently checks the population bound and its rectangle-projection formula before simulation.

\clearpage
\section{Figures}
\end{document}